\documentclass[9pt]{article}
\usepackage[top = 1in, bottom = 1in, left = 1in, right = 1in]{geometry}
\usepackage{graphicx} 
\usepackage{amsmath}
\usepackage[sort]{natbib}
\usepackage{threeparttable}
\usepackage{amssymb}

\usepackage{extarrows}

\usepackage[utf8]{inputenc}
\usepackage[english]{babel}

\usepackage{setspace}
\usepackage{multicol}

\usepackage{lipsum}
\usepackage{threeparttable}

\usepackage{amsthm}

\theoremstyle{definition}

\usepackage{marginnote}

\usepackage{parskip}                
\usepackage{subcaption}             

\usepackage[symbol]{footmisc}

\usepackage{color}
\usepackage{soul}  

\usepackage{makecell}

\usepackage[colorinlistoftodos]{todonotes}

\usepackage{xcolor}
\usepackage[most]{tcolorbox}

\usepackage{afterpage}

\usepackage{array,multirow,graphicx}
\usepackage{float}

\usepackage{color, colortbl}
\usepackage[first=0,last=9]{lcg}

\definecolor{Gray}{gray}{0.9}
\definecolor{LightCyan}{rgb}{0.88,1,1}

\usepackage{footnote}

\usepackage{ragged2e} 

\usepackage{lineno}

\usepackage[useregional]{datetime2}

\usepackage{hyperref}
\hypersetup{
    colorlinks=true,
    linkcolor=blue,
    citecolor=blue,
    filecolor=magenta,      
    urlcolor=cyan,
    pdftitle={Overleaf Example},
    pdfpagemode=FullScreen,
    }

\usepackage{xcolor}

\colorlet{lightgreen}{green!10}

\newcolumntype{M}[1]{>{\centering\arraybackslash}m{#1}}

\usepackage{authblk}

\title{piCurve: an R package for modeling photosynthesis–irradiance curves}
\author[1]{Mohammad M. Amirian\footnote{Corresponding Author: m.amirianmatlob@dal.ca}}
\author[1]{Andrew J. Irwin}
\affil[1]{Department of Mathematics and Statistics, Dalhousie University, Halifax NS}

\begin{document}
\maketitle
\begin{abstract}

Photosynthesis–irradiance (P–I) curves are foundational for quantifying primary production, parameterizing ecosystem and biogeochemical models, and interpreting physiological acclimation to light. Despite their broad use, researchers lack a unified, reproducible toolkit to fit, compare, and diagnose the many PI formulations that have accumulated over the last century. We introduce piCurve, an R package that standardizes the modeling of P–I relationships, with a library of widely used light-limited, light-saturated, and photoinhibited formulations and a consistent statistical framework for estimation and comparison. With the total of 24 P-I models, piCurve supports mean squared error (MSE) and maximum likelihood estimation (MLE), provides uncertainty quantification via information matrix (Hessian), and includes automated, data-informed initialization to improve convergence. Utilities classify P–I data into light-limited, light-saturated, and photoinhibited regions, while plotting and 'tidy' helpers streamline workflow and reporting. Together, these features enable reproducible analyses and fair model comparisons, including for curves exhibiting a plateau followed by photoinhibition.
\\
\\

\textbf{Availability}: R package on GitHub (\hyperlink{https://github.com/Mohammad-Amirian/piCurve}{https://github.com/Mohammad-Amirian/piCurve}). Example data (eight P–I incubations) and vignettes are included. 
\end{abstract}

\section{Introduction}
\label{sec:intro}

The photosynthetic response to light is often summarized by the photosynthesis–irradiance (P–I) curve, with three characteristic regions: light-limited, light-saturated, and photoinhibited, parameterized by the initial slope ($\alpha$), the maximum photosynthetic rate (P$_{\max}$), dark respiration rate (R) and a photoinhibition parameter ($\beta$). These parameters are widely used to compute water-column–integrated primary production \cite{devred2025net} and to underpin ecological \cite{sloughter2019seasonal} and biogeochemical models \cite{follows2007emergent}. However, the literature contains many distinct mathematical P–I formulations (Tables. 1 \& S2 of \cite{m2025parameterization}), developed independently over time. In practice, analysts face several recurring obstacles: (i) disparate model parameterizations, where models differ in algebraic form, symbols, and parameter definitions; (ii) uneven or ad hoc optimization procedures, often relying on nonlinear algorithms that are sensitive to starting values, convergence criteria, parameter bounds, and data points; (iii) imbalanced data coverage across light levels, which can bias parameter estimation; (iv) subjective handling of photoinhibition, including manual removal of high-irradiance points; and (v) limited reproducibility across studies, where methodological differences prevent fair comparisons even when similar datasets are analyzed. A unified, transparent framework is needed to fit multiple models consistently, quantify uncertainty, and support reproducible, apples-to-apples comparisons across datasets and model families.
\section{Knowledge Gap}
\label{sec:Knowledge g}

Prior to recent advances that explicitly capture the plateau preceding photoinhibition, researchers often truncated or excluded high-irradiance observations to avoid convergence issues or pathological fits, introducing subjectivity and bias. While individual R scripts implement specific formulations, there has been no comprehensive, extensible, open-source library that: (a) gathers commonly used P–I models in one place; (b) standardizes fitting criteria, initialization routines, and diagnostics across both balanced and unbalanced datasets; and (c) includes dedicated tools for photoinhibited curves. The absence of such a framework means that differences in data structure, optimization settings, and model parameterization can all lead to inconsistent or irreproducible results. The \texttt{piCurve} package addresses this gap by unifying over a century of P–I model development into a single, coherent API and analysis pipeline, enabling robust, reproducible, and scalable applications in ecological and biogeochemical research. 
\section{Modeling Functions}
\label{sec:modeling-fn}

A photosynthesis–irradiance (P–I) curve typically consists of three distinct regions: \textit{light-limited}, \textit{light-saturated}, and \textit{photoinhibited}. To capture this full range of responses, \texttt{piCurve} implements 24 independent models: 1 light-limited model, 7 light-saturated models (Table. \ref{tb:ls-eqs}), and 16 photoinhibition models (Table. \ref{tb:ph-eqs}).

The light-limited model is a simple linear regression describing phytoplankton photosynthetic response under low irradiance, where photosynthesis increases proportionally with light. The light-saturated models encompass a variety of well-established algebraic forms—including piecewise linear \cite{blackman1905optima}, rectangular hyperbola \cite{baly1935kinetics}, exponential \cite{webb1974carbon}, hyperbolic tangent \cite{jassby1976mathematical}, non-rectangular hyperbola \cite{prioul1977partitioning}, and generalized rectangular hyperbola \cite{bannister1979quantitative}, each capable of representing the plateau region where photosynthesis approaches its maximum rate. The photoinhibition models extend these light-saturated forms by multiplication by an exponential decay \cite{steele1962environmental, platt1981photoinhibition, neale1987photoinhibition},  term or a saturating function of the reciprocal of irradiance \cite{m2025parameterization}, allowing accurate representation of the decline in photosynthesis at very high irradiance.

In all cases, an optional dark respiration term can be included, enabling estimation of net or gross photosynthetic rates \cite{sloughter2019seasonal} according to user requirements. This broad suite of formulations ensures that \texttt{piCurve} can flexibly model virtually any observed P–I curve shape.

The recommended formulation for fitting a PI curve is the double-tanh model (Eq.~\ref{eq:2tanh}), as it accurately captures the plateau phase in the presence of photoinhibition. Unlike other approaches, this model estimates all parameters directly from the data, avoiding artificial variability in $P_{\max}$ often introduced by intermediate calculations. This direct estimation improves interpretability, particularly for the photoinhibition rate $\beta$, and allows a straightforward computation of the irradiance at which photoinhibition reaches its maximum $I_\beta = P_{\max} / \beta$. In contrast, other models (Ph01–Ph08; Table.~\ref{tb:ph-eqs}) require first estimating the theoretical maximum photosynthesis rate $P_s$ and then numerically deriving $P_{\max}$ from multiple fitted parameters. This two-step process can amplify uncertainty, especially in unbalanced datasets, as error in any single parameter propagates to $P_{\max}$ \cite{m2025parameterization}.
\begin{equation}
\label{eq:2tanh}
     P = 
    P_{\max} \tanh \left( \dfrac{I}{I_\alpha} \right) 
     \tanh{
     \left[ \left( \dfrac{I_\beta}{I} \right)^{\gamma}
     \right]
     }, 
     \hspace{1cm}
     \text{s.t}
     \hspace{0.5cm}
     \gamma = \cosh^2(1) \approx 2.38
\end{equation}
Here, $\alpha$ denotes the photosynthetic efficiency at low irradiance. When $\beta = 0$, the equation (\ref{eq:2tanh}) simplifies to the Jassby model (LS5, Table~\ref{tb:ls-eqs}). In \texttt{piCurve}, LS5 is the default setting when photoinhibition is absent; if photoinhibition is detected, the package automatically switches to the double-tanh formulation (Ph10, Table~\ref{tb:ph-eqs}). All the available models are listed in Tables. \ref{tb:ls-eqs} \& \ref{tb:ph-eqs}.
\section{Package Overview}

The \texttt{piCurve} package was developed as a direct outcome of our recent work \cite{m2025parameterization}, in which we compiled and compared a suite of existing and newly developed P–I models using $\sim$ 4,000 independent open-ocean field P-I incubation data samples. During that project, we found fair model comparison requires fitting all candidate models with the same statistical optimization methods and evaluation criteria, ensuring differences in performance reflect the models themselves rather than inconsistencies in the fitting approach.
This need is amplified by common characteristics of field P–I datasets: unbalanced sampling, with as few as 8 and as many as 60 irradiance levels per experiment; strong nonlinearity in model forms, making convergence sensitive to initialization and optimizer settings; parameter interdependence ($\alpha$, P$_{\max}$, $\beta$, and R are often correlated), meaning that small sample sizes can inflate uncertainty and bias estimates; inherent noise in in-situ measurements; and frequent photoinhibition, for which there has been no widely adopted, standardized modeling protocol.

Including models from 1905 to 2025, the \texttt{piCurve} package offers a robust, reproducible, and extensible R toolkit in which the models are validated against $\sim$ 4,000 field experiments and $>$ 15,000 simulated P–I datasets. 
Models that did not pass these validation tests include the logarithmic models \cite{talling1957photosynthetic, chalker1980modeling} (for details, see \cite{m2025parameterization}). Currently, \texttt{piCurve} provides 14 functions, organized into three categories (Table. \ref{tbl:avail_fn}):

\begin{itemize}
    \item \textbf{Core modeling \& fitting} — Functions for model definition, automated initialization, and fitting using consistent statistical criteria, and dataset classifier,
	
    \item \textbf{Inference \& diagnostics} — Functions for model adequacy testing, goodness-of-fit metrics, uncertainty estimation, and information criteria,
	\item \textbf{Utilities} (data, prediction, plotting) — Functions for validating input data, generating predictions on a fine irradiance grid, adding confidence intervals, tidying output, and producing publication-quality figures.

\end{itemize}
\section{Workflow \& Usage}

The functions in \texttt{piCurve} are designed for ease of use, enabling rapid, reliable, and reproducible analysis of photosynthesis–irradiance data. A complete list of functions is provided in Table 1; here, we focus on three core functions that form the central workflow of the package.

\subsection{Model fitting}
\begin{figure}
    \centering
    \includegraphics[width=0.7\linewidth]{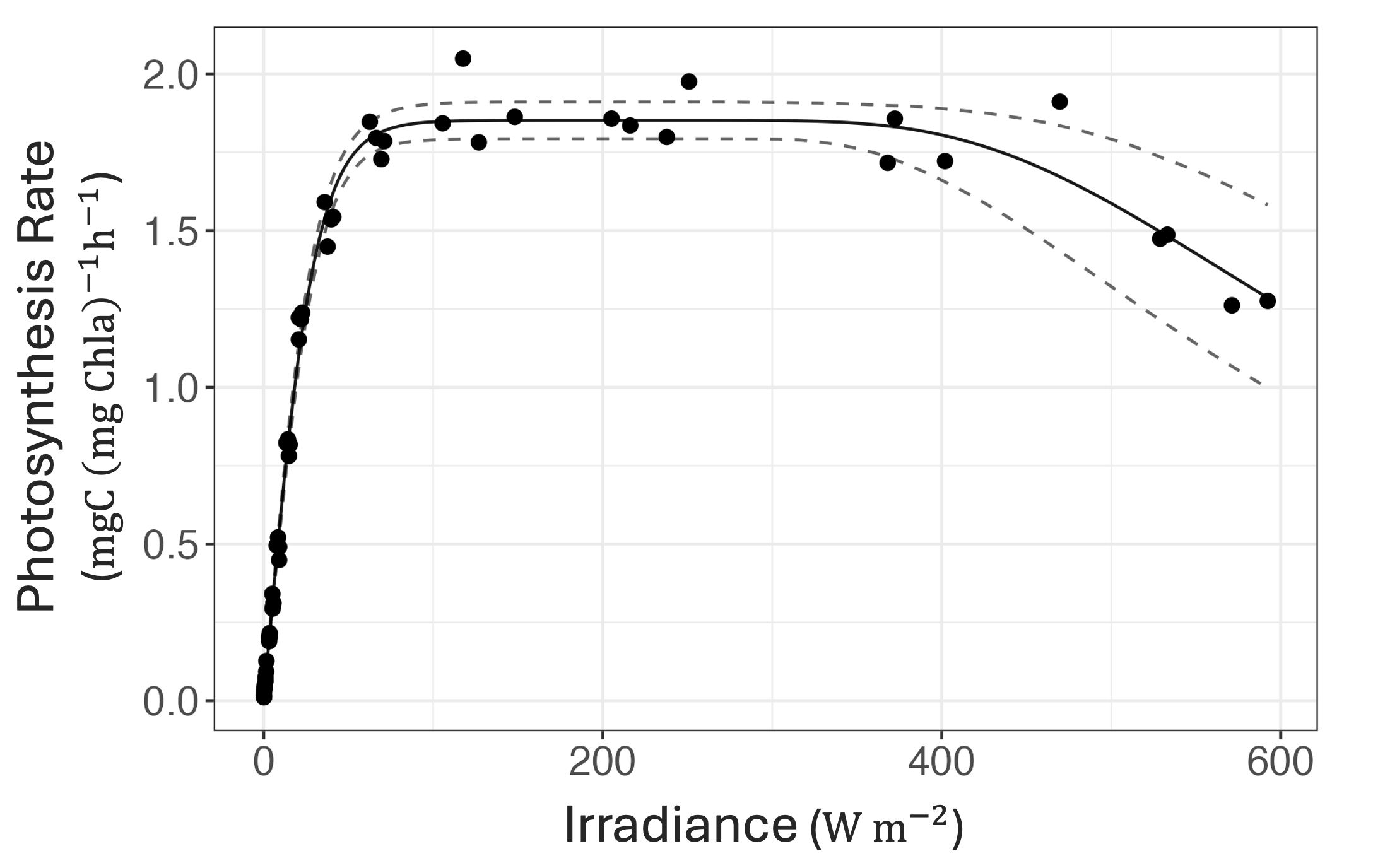}
  \caption{
  Fit of the double-tanh model (Eq.~\ref{eq:2tanh}; solid line) to a representative PI dataset (dots). Dashed lines indicate 95\% prediction confidence intervals. Data are from the \texttt{piCurve} library (\texttt{pi\_information} = PI002413).
  }
    \label{fig:2tanh-fit}
\end{figure}

The \texttt{piCurve} workflow is straightforward: users provide their dataset to the `\texttt{Fit\_piModel}' function for model fitting and select a statistical approach (default: MSE). The resulting fit can be organized into a tidy format using `\texttt{Tidy\_piCurve}', and visualized with `\texttt{Plot\_piCurve}' (Fig.~\ref{fig:2tanh-fit}). This combination of functions enables users to go from raw data to publication-ready figures in a single, reproducible workflow.

\subsection{Model comparison}
\begin{figure}
    \centering
    \includegraphics[width=1\linewidth]{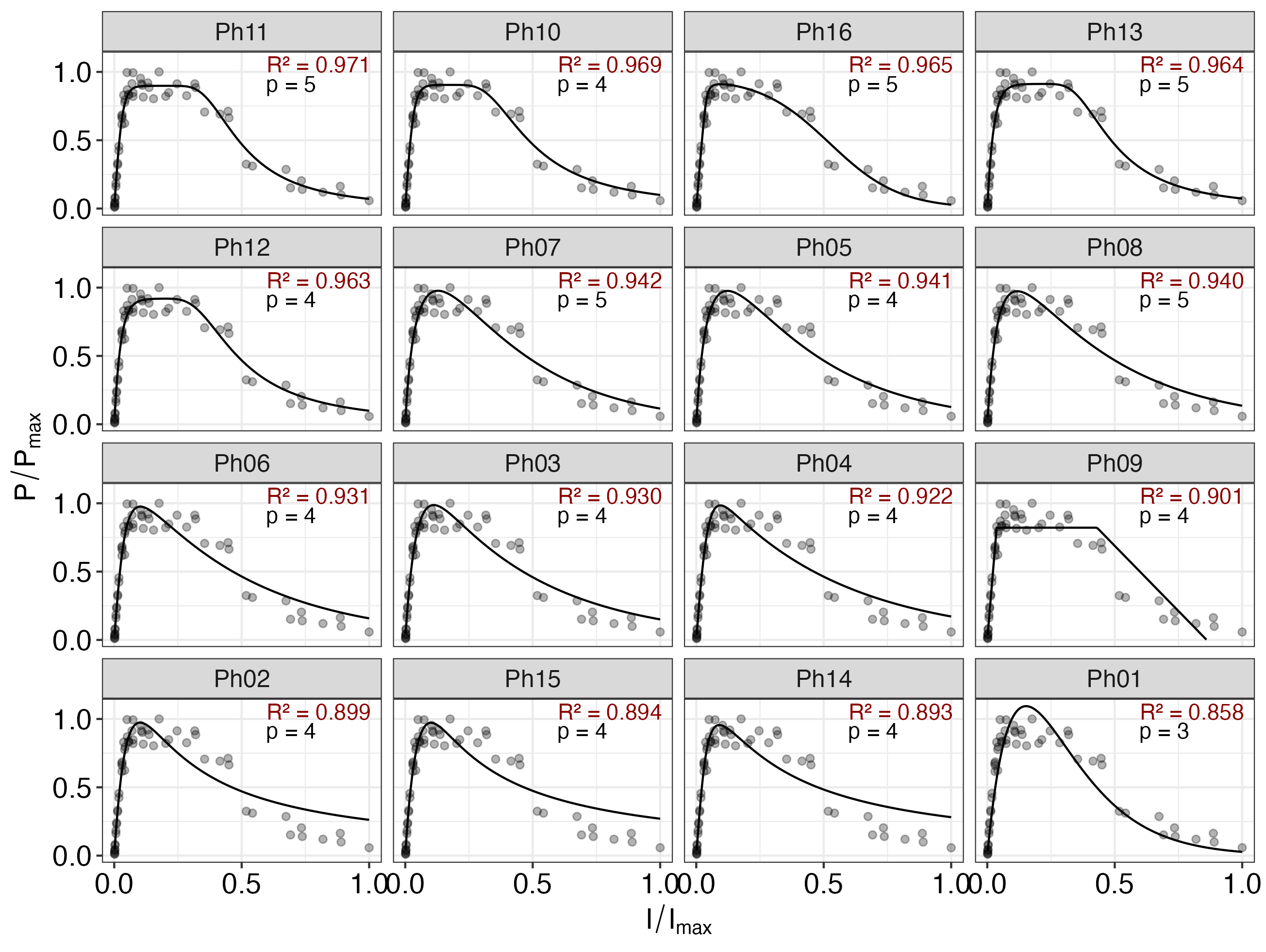}
 \caption{
    Fits of all photoinhibition models in \texttt{piCurve} (black lines) to a representative P–I dataset (gray dots). $R^2$ indicates the variance explained by each model, and $p$ denotes the number of model parameters. Axes are normalized by their respective maximum values. Panels are sorted from highest to lowest $R^2$. Data are from the \texttt{piCurve} library (\texttt{pi\_information} = PI002413). Equations are given in Table.~\ref{tb:ph-eqs}.
}
    \label{fig:fit-allModels}
\end{figure}

The \texttt{piCurve} package allows users to systematically compare all available models. To do so, select a model of interest (see the \texttt{Name} column in Tables~\ref{tb:ls-eqs} and \ref{tb:ph-eqs}) and specify it in the \texttt{model\_name} argument of the \texttt{Fit\_piModel} function. By iterating over the list of candidate models, each model can be fitted to the dataset using consistent optimization criteria, enabling a fair, side-by-side performance evaluation (Fig.~\ref{fig:fit-allModels}).

\subsection{Data classification}
\begin{figure}
    \centering
    \includegraphics[width=1\linewidth]{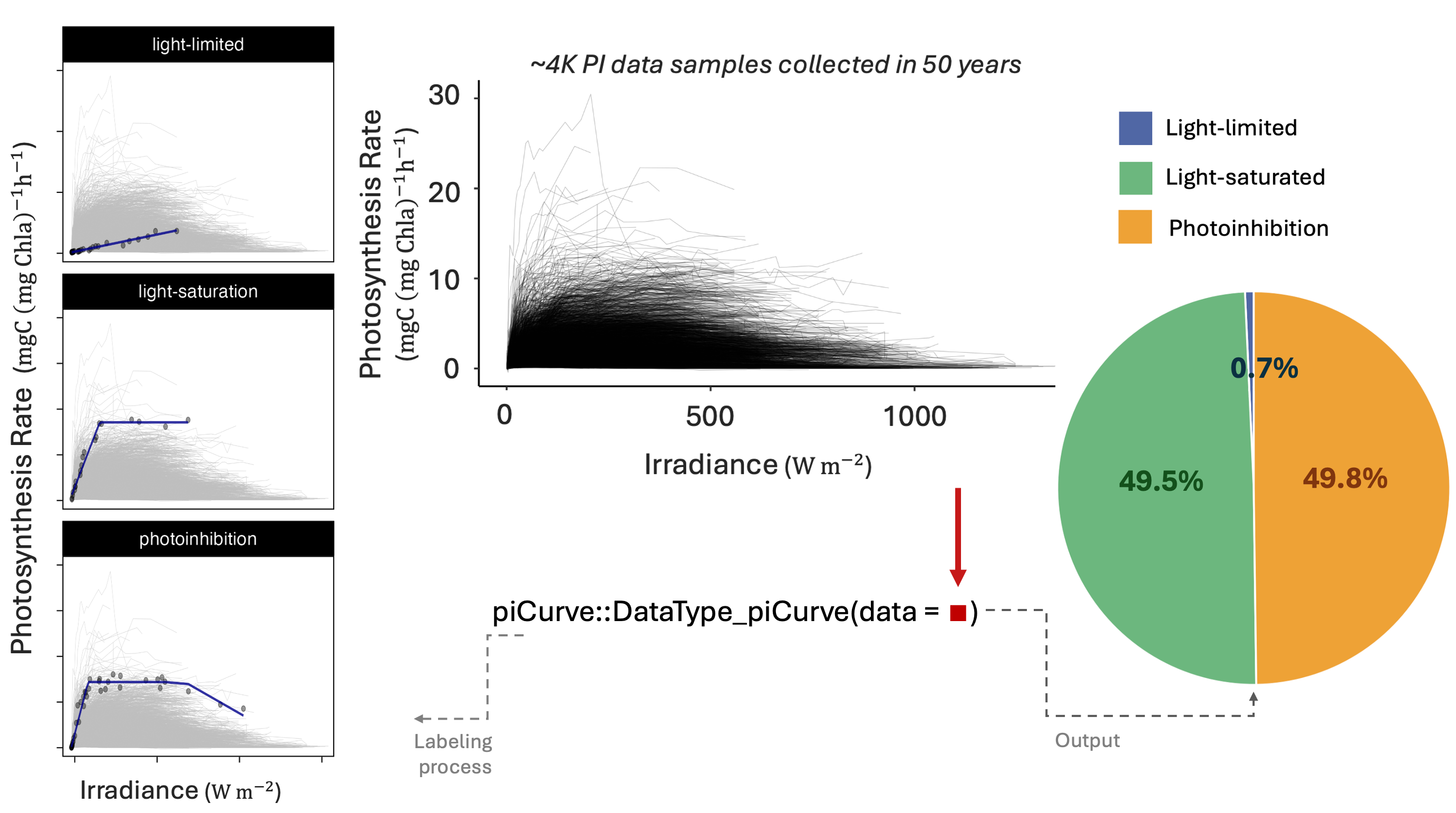}
    \caption{Labeling the PI curve metadata using the \texttt{DataType\_piCurve} function, categorized by photoinhibition, light-saturation, and light-limited regions.}

    \label{fig:data-classify}
\end{figure}

The \texttt{piCurve} package includes a built-in classifier, `\texttt{DataType\_piCurve}', for labeling individual P–I datasets or entire databases according to the overall light–photosynthesis relationship. Users simply provide the dataset to `\texttt{DataType\_piCurve}', which automatically categorizes each curve as light-limited, light-saturated, or photoinhibited (Tables.~\ref{tb:ls-eqs} \& \ref{tb:ph-eqs}). This function can be applied to both individual experiments and large-scale databases, enabling consistent classification across diverse studies.
As an example, we applied this classifier to the $\sim$ 4,000 open-ocean P–I incubation datasets analyzed in our recent study \cite{m2025parameterization}, generating a relative frequency  of curve types across the dataset (Fig.~\ref{fig:data-classify}, right panel). 

\section{Discussion}
\label{sec:dis}


The \texttt{piCurve} package fills a long-standing need by providing a standardized, reproducible, and extensible framework for modeling photosynthesis–irradiance (P–I) relationships. By integrating 24 widely used light-limited, light-saturated, and photoinhibition models within a unified statistical workflow, it allows researchers to compare model performance on an equal footing, removing methodological inconsistencies that have historically limited cross-study synthesis.

One of the key contributions of \texttt{piCurve} is its explicit handling of photoinhibition, with dedicated formulations for curves that exhibit a distinct plateau before declining. A recent analysis shows that roughly 75\% of photoinhibited datasets fall into this plateau category. Capturing this feature not only improves the realism of fits for high-irradiance datasets but also eliminates the need for subjective data truncation, a common practice in earlier studies. The built-in curve classifier, \texttt{DataType\_piCurve,} further supports large-scale ecological assessments by rapidly categorizing curves as light-limited, light-saturated, or photoinhibited, ensuring consistent labeling across diverse datasets.

By combining model fitting, curve classification, and visualization into a single, coherent workflow, \texttt{piCurve} supports both exploratory analyses and high-throughput processing of large datasets. The reproducible framework aligns with open-science principles, and the open availability of the package and example datasets via public repositories ensures that results are transparent, verifiable, and easily extended by the community.

In summary, \texttt{piCurve} establishes a robust foundation for advancing both methodological rigor and ecological insight in P–I research. Its open-source design and modular architecture invite community contributions, ensuring that it can evolve to meet emerging challenges in marine primary production modeling and beyond.

\pagebreak
\begin{table}[ht]
\centering
\caption{Available functions in \texttt{piCurve} library. Functions marked with an asterisk ($^\star$) support multi-core options to accelerate high-throughput analyses.}
\begin{tabular}{p{0.25\linewidth} p{0.68\linewidth}}
\textbf{Function} & \textbf{Description} 
\\
\hline
& \\
[-1ex]
\multicolumn{2}{l}{\textit{Core modeling \& fitting}} 
\\
[1.5ex]
\rowcolor{gray!20}
\texttt{Model\_piCurve()} & 
Define or select the P–I model formulation. 
\\
[1.5ex]
\texttt{Fit\_piModel()} & 
Fit P–I models (MSE or MLE), with options for Hessian, bounds, and optimizer technique (default: Nelder–Mead). 
\\
[4.5ex]
\rowcolor{gray!20}
\texttt{get\_start\_piPars()} &
Suggest informed starting values for nonlinear optimization. 
\\
[1.5ex]
$^\star$\texttt{DataType\_piCurve()} &
Automatically classify dataset as light-limited, light-saturated, or photoinhibited. Includes built-in example dataset (\texttt{piDataSet}) with eight independent P–I incubations for testing and demonstration. 
\\
[6.5ex]
\hline
& \\
[-1ex]
\multicolumn{2}{l}{\textit{Inference \& diagnostics}} 
\\
[1.5ex]
\rowcolor{gray!20}
\texttt{AIC\_AICc\_BIC\_piCurve()} &
Compute information criteria for model selection. 
\\
[1.5ex]
\texttt{R2\_piCurve()} &
Compute goodness-of-fit via $R^2$ and adjusted $R^2$. 
\\
[1.5ex]
\rowcolor{gray!20}
\texttt{MSE\_piCurve()} & 
Compute mean squared error (fit adequacy / loss). 
\\
[1.5ex]
\texttt{InfoMat\_piCurve()} &
Estimate the observed information matrix when \texttt{Hessian = TRUE} is not provided to \texttt{Fit\_piModel()}. 
\\
[4.5ex]
\rowcolor{gray!20}
\texttt{ConfInt\_piCurve()} &
Calculate confidence intervals for fitted parameters. 
\\
[1.5ex]
\texttt{ReCal\_CI\_piCurve()} & 
Recalculate or adjust confidence intervals. 
\\
[1.5ex]
\hline
& \\
[-1ex]
\multicolumn{2}{l}{\textit{Utilities (data, prediction, plotting)}} 
\\
[1.5ex]
\rowcolor{gray!20}
\texttt{FormatCheck\_piCurve()} & 
Validate input P–I data structure. 
\\
[1.5ex]
\texttt{highRes\_piPred()} &
Generate high-resolution predictions over irradiance. 
\\
[1.5ex]
\rowcolor{gray!20}
$^\star$\texttt{addCI\_to\_piPred()} &
Attach prediction/interval bands to predictions. 
\\
[1.5ex]
\texttt{Tidy\_piCurve()} &
Tidy \texttt{Fit\_piModel()} output into analysis-ready tables of parameter estimates and fit statistics. 
\\
[4.5ex]
\rowcolor{gray!20}
$^\star$\texttt{Plot\_piCurve()} & Create publication-ready plots of data, fit, and optional confidence bands. 
\\
[1.5ex]
\hline
\end{tabular}
\label{tbl:avail_fn}
\end{table}

\clearpage

\bibliographystyle{plain}
\bibliography{refs.bib}

\pagebreak
\section*{Appendix}

\setcounter{equation}{0}
\setcounter{figure}{0}
\setcounter{table}{0}
\renewcommand{\theequation}{A\arabic{equation}}
\renewcommand{\thefigure}{A\arabic{figure}}
\renewcommand{\thetable}{A\arabic{table}}
\renewcommand{\bibnumfmt}[1]{[A#1]}

\renewcommand{\arraystretch}{2} 

\begin{table}[htbp]
\centering
\caption{
Models used to formulate light-limited and light-saturating PI curves, parameterized by $P_{\max}$, $I_{\alpha} = P_{\max}/\alpha$, and shape parameters $\theta$ and $\gamma$ (with $0 < \theta < 1$). The \textit{Name} column indicates the identifier used to call each model in the \texttt{piCurve} package. All models include an optional constant intercept $R$ representing dark respiration, which is omitted from the table for brevity.
}

\begin{tabular}{lll}
\textbf{Name} & \textbf{Equation} & 
\textbf{References} \\
\hline
\rowcolor{gray!20} \multicolumn{3}{l}{\textit{Light-limited models}} 
\\
lm & $ \alpha I$ & Linear Regression \cite{blackman1905optima}
\\ [0.2cm]
\rowcolor{gray!20} \multicolumn{3}{l}{\textit{Light-saturating models}} 
\\
LS1 & $P_{\max}[{I+I_\alpha - |I-I_\alpha|}]/{2I_\alpha}$ & Blackman 1905 \cite{blackman1905optima} 
\\
LS2 & $P_{\max}{I}/({I + I_\alpha})$ & Baly 1935 \cite{baly1935kinetics}
\\
LS3 & $P_{\max}{I}/{\sqrt{I^2 + I^2_\alpha}}$ & Smith 1936 \cite{smith1936photosynthesis}
\\
LS4 & $P_{\max}(1-e^{-I/I_\alpha})$ & Webb et al. 1974 \cite{webb1974carbon}
\\
LS5 & $P_{\max} \tanh \left( {I}/{I_\alpha} \right)$ & Jassby et al. 1976 \cite{jassby1976mathematical}
\\
LS6 & ${P_{\max}}
\left[(I/I_\alpha + 1) -\sqrt{(I/I_\alpha + 1)^2 - 4 \theta
(I/I_\alpha)} \right]/{2 \theta}$ & Prioul et al. 1977 \cite{prioul1977partitioning}
\\
LS7 & $P_{\max} {I}/{(I^\gamma+I^\gamma_\alpha)^{1/\gamma}}$ & Bannister 1979 \cite{bannister1979quantitative}
\\[0.2cm]
\hline
\end{tabular}
\label{tb:ls-eqs}
\end{table}

\renewcommand{\arraystretch}{2.5} 

\begin{table}[htbp]
\centering
\caption{
Models used to formulate photoinhibition PI curves, parameterized by $P_{\max}$, $P_{s}$, $I_{\alpha} = P_{\max}/\alpha$, $I^s_{\alpha} = P_{s}/\alpha$, $I_{\beta} = P_{\max}/\alpha$, $I^s_{\beta} = P_{s}/\beta$ and shape parameters $\theta$ and $\gamma$ (with $0 < \theta < 1$). The \textit{Name} column indicates the identifier used to call each model in the \texttt{piCurve} package. All models include an optional constant intercept $R$ representing dark respiration, which is omitted from the table for brevity.
}
\begin{tabular}{lll}
\textbf{Name} & \textbf{Equation} & 
\textbf{References} \\
\hline
Ph01 & $P_s~ (I/I_{\alpha}^s) ~e^{1-( I/ I_{\alpha}^s)}$ & Steele 1962 \cite{steele1962environmental}
\\
Ph02 & $P_s ~ {(I/I_{\alpha}^s)} / [{ ~ I^2 / (I_{\alpha}^s I_{\beta}^s) + (I/I_{\alpha}^s) + 1}]$ & Peeters et al. 1978 \cite{peeters1978relationship}
\\
Ph03 & $P_s~ (1-e^{-I/I_{\alpha}^s})e^{-I/I_{\beta}^s}$ & Platt et al. 1981 \cite{platt1981photoinhibition}
\\
Ph04 & $P_s~ \tanh \left(I/I_{\alpha}^s\right) ~ e^{-I/I_{\beta}^s}$ &  Neale et al. 1987 \cite{neale1987photoinhibition}
\\
Ph05 & $ P_s~ {I}/({I + I_{\alpha}^s})  ~ e^{-I/I_{\beta}^s}$ &  
Amirian et al. 2025 \cite{m2025parameterization}
\\
Ph06 & $ P_s~ {I} / {\sqrt{I^2 + (I_{\alpha}^s)^2}}  ~ e^{-I/I_{\beta}^s}$ & 
Amirian et al. 2025 \cite{m2025parameterization}
\\
Ph07 & $ P_s~ {I} / {(I + (I_{\alpha}^s))^{1/b}} ~ e^{-I/I_{\beta}^s}$ &  
Amirian et al. 2025 \cite{m2025parameterization}
\\
Ph08 & $\frac{P_{s}}{2 \theta} 
\left[ 
(I/I_\alpha + 1) -
\sqrt{(I/I_\alpha + 1)^2 - 4 \theta(I/I_\alpha)} \right] ~ e^{-I/I_{\beta}^s}$ &  
Amirian et al. 2025 \cite{m2025parameterization}
\\
Ph09 & $ 
\begin{cases} 
\alpha I & I \leq P_{\max} / \alpha \\ 
P_{\max} & P_{\max} / \alpha < I \leq P_{\max} / \beta\\ 
-\beta I & I > P_{\max} / \beta 
\end{cases}$ & 
Amirian et al. 2025 \cite{m2025parameterization}
\\
Ph10 & $ P_{\max} \tanh \left( {I}/{I_{\alpha}} \right) \tanh{ \left[ \left( {I_{\beta}}/{I} \right)^{\cosh^2(1)} \right] }$ & Amirian et al. 2025 \cite{m2025parameterization}
\\
Ph11 & $ P_{\max} \tanh \left( {I}/{I_{\alpha}} \right) \tanh{ \left[ \left( {I_{\beta}} / {I} \right)^{\gamma} \right] }$ & Amirian et al. 2025 \cite{m2025parameterization} 
\\
Ph12 & $ P_{\max}  \left[1 - \exp{\left( -{I}/{I_{\alpha}} \right)} \right] \tanh{ \left[ \left( {I_{\beta}}/{I} \right)^{\cosh^2(1)} \right] }$ & Amirian et al. 2025 \cite{m2025parameterization}
\\
Ph13 & $ P_{\max} \left[1 - \exp{\left( -{I}/{I_{\alpha}} \right)} \right] \tanh{ \left[ \left( {I_{\beta}}/{I} \right)^{\gamma} \right] }$ & Amirian et al. 2025 \cite{m2025parameterization}
\\
Ph14 & $ P_{\max}  \tanh{\left({I}/{I_{\alpha}}\right)} \left[1 - \exp{\left(-{I_{\beta}}{/I}\right)} \right]$ & Amirian et al. 2025 \cite{m2025parameterization}
\\
Ph15 & $ P_{\max} \left[1 - \exp{\left(-{I}/{I_{\alpha}}\right)} \right] \left[1 - \exp{\left(-{I_{\beta}}/{I}\right)} \right]$  & Amirian et al. 2025 \cite{m2025parameterization}
\\
Ph16 & 
$ \frac{P_{\max}}{2 \theta} 
\left[1 + \theta_\beta ~ (I / I_\alpha) -
\sqrt{ \theta_\beta ~  (I / I_\alpha)^2 - 4 \theta ~ (I / I_\alpha) + 1} \right]$ & Fasham et al. 1983 \cite{fasham1983photosynthetic}
\\

\hline
\end{tabular}
\label{tb:ph-eqs}
\end{table}

\end{document}